\documentclass[prl,aps,showpacs,superscriptaddress,twocolumn]{revtex4}
\usepackage{color} %used for font color    
\usepackage{times}
\usepackage{amssymb} %maths            
\usepackage{amsmath} %maths                      
\usepackage{ragged2e}
\usepackage{graphicx}
\usepackage{epstopdf}
\usepackage[english]{babel}
\usepackage{mathrsfs}
\usepackage{textcomp}
\usepackage{amsthm}
\usepackage{amsfonts}
\usepackage{soul}
\usepackage{braket} 

\newcommand{\be}{\begin{equation}}
\newcommand{\ee}{\end{equation}}
\newcommand{\bea}{\begin{eqnarray}}
\newcommand{\eea}{\end{eqnarray}}
\newcommand{\ba}{\begin{array}}
\newcommand{\ea}{\end{array}}

\newcommand{\change}[1]{\textcolor{black}{#1}}

\begin{document}

\title{Haldane Topological Orders in %the Deformed 
Motzkin Spin Chains}
\author{L. Barbiero}
\affiliation{CNR-IOM DEMOCRITOS Simulation Center, Via Bonomea 265, I-34136
Trieste, Italy}
\affiliation{Center for Nonlinear Phenomena and Complex Systems, UniversitŽ Libre de Bruxelles, CP 231, Campus Plaine, B-1050 Brussels, Belgium}
%\affiliation{Dipartimento di Fisica e Astronomia ``G. Galilei'', 
%Universit\`a di Padova, via F. Marzolo 8, I-35131, Padova, Italy}
\author{L. Dell'Anna}
\affiliation{Dipartimento di Fisica e Astronomia ``G. Galilei'', 
Universit\`a di Padova, via F. Marzolo 8, I-35131, Padova, Italy}
\author{A. Trombettoni}
\affiliation{CNR-IOM DEMOCRITOS Simulation Center, Via Bonomea 265, I-34136
Trieste, Italy}
\affiliation{SISSA and INFN, Sezione di Trieste, Via Bonomea 265, I-34136 
Trieste, Italy}
\author{V. E. Korepin}
\affiliation{C. N. Yang Institute for Theoretical Physics, 
Stony Brook University, NY 11794, USA}

\begin{abstract}
Motzkin spin chains are frustration-free models whose ground-state is a 
combination of Motzkin paths. The weight of such path contributions 
can be controlled by a deformation parameter $t$. As a function of the latter 
these models, beside the formation of domain wall structures, exhibit 
gapped Haldane topological orders with constant decay of the string order parameters for $t<1$. 
A behavior compatible with a Berezinskii-Kosterlitz-Thouless phase transition at $t=1$ is also presented.
By means of numerical calculations we show 
that the topological properties of the Haldane phases 
depend on the spin 
value. This allows to classify different kinds of hidden antiferromagnetic 
Haldane gapped regimes associated to nontrivial features like 
symmetry-protected topological order. 
Our results from one side allow to clarify the physical 
properties of Motzkin frustration-free chains and from 
the other suggest them as a new 
interesting and paradigmatic class of local spin Hamiltonians.
\end{abstract}

\maketitle

Spin chains play a crucial role in many fundamental 
physical phenomena like magnetism \cite{auerbach}, 
quantum phase transitions \cite{sachdev}, topological orders \cite{wen} 
and quantum computation \cite{nielsen}. A fundamental contribute 
to the understanding of spin chains is provided by the 
the seminal papers by Haldane \cite{haldane} 
where a new topological phase, the Haldane phase (HP), uniquely detectable via a non-local string order parameter (SOP) \cite{denNijs} has been discovered 
for spin-$1$ XXZ Heisenberg chains.
This has driven significant efforts 
to look for new kinds of models whose topological order 
can be described in terms of a SOP \cite{wen1} 
motivating the discovery of the celebrated Affleck-Kennedy-Lieb-Tasaki (AKLT) 
model \cite{aklt}. Although the argument of Haldane is given for integer spin 
chains, only integer spin XXZ-like and AKLT-like chains own topological HP 
and is therefore non-trivial to find and study new classes of Hamiltonians 
where HP emerges.
\begin{figure}[!]
\includegraphics[height=8.5cm,width=\columnwidth]{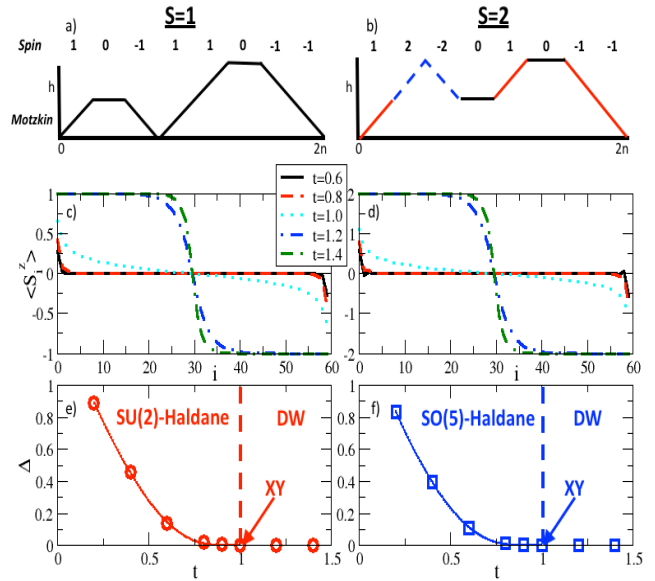}
\caption{(Color online) \textit{Upper panels}: Cartoons of a 
possible Motzkin path and its representation in terms of spins for the 
two cases \textit{a)} uncolored $s=1$ and \textit{b)} colored $s=2$. 
\textit{Central panels}: DMRG local magnetization 
for a system of length $2n=60$ at different $t$ deformation values
$\langle S^z_i\rangle$ for \textit{c)} $s=1$ and \textit{d)} $s=2$. 
\textit{Lower panels}: Thermodynamic limit of the gap $\Delta=E_1-E_0$ 
as a function of $t$ for \textit{e)} $s=1$ (red circles) 
and \textit{f)} $s=2$ (blue squares). The continuos lines are 
fitted with the form $\sim \exp({-b/\sqrt{t_c-t}})$ 
with $t_c=1$ and $b$ a fitting parameter. 
The thermodynamic limit 
is extrapolated by using chains of length up to $2n=60$. 
All the DMRG simulation are performed by keeping at most $1024$ 
DMRG states and $5$ finite size sweeps with an error energy 
$<10^{-9}$ $(10^{-7})$ for $s=1$ ($s=2$).}
\label{fig1}
\end{figure}
Thanks to the strongest quantum "resource", namely the entanglement, spin models have also a fundamental role 
in the simulation of quantum logical gates 
for quantum computation \cite{nielsen}. 
For this reason finding and studying Hamiltonians 
with highly entangled spins 
is currently one of the most challenging and intriguing fields in 
quantum physics.\\
In this direction local integer frustration-free spin Hamiltonians 
whose ground-state can be expressed as a combination 
of Motzkin paths \cite{book} have been recently 
introduced \cite{bravyi12,movassagh14}. Among others interesting aspects, 
their importance is given by the fact that they own a level of 
entanglement entropy which strongly exceeds the one exhibited by 
other previously known local models. Relevantly, 
also for half-integer spins, a similar class of Hamiltonians, the Fredkin spin chains, exhibiting the same features 
\cite{dellanna,olof} has been introduced. 
In addition to their entanglement properties 
Motzkin chains own further very peculiar properties. 
Indeed, even if they are purely local models, for high spin values $s$  
(i.e., $s\ge 2$) they behave as \textit{de facto} 
long range Hamiltonians being able to violate cluster 
decomposition properties (CDP) and area law (AL) scaling of the entanglement 
entropy \cite{dellanna}. 
%and light cone-like excitation propagation \cite{dellanna}. 
Very recently, a deformed version of Motzkin \cite{klich} 
and of Fredkin \cite{olof2} chains have been introduced, and their gap 
studied \cite{ramis}, with the contribution of Motzkin or Fredkin 
paths to the ground-state being weighted through the introduction 
of a parameter $t$.\\
Due to the aforementioned arguments it appears clear as these new models 
are both very interesting by themselves and they could open the path 
towards fundamental applications. This motivates us 
to investigate a Motzkin chain for different spin 
values and in presence of path deformations. Here after an 
introduction of the model in terms of deformed Motzkin paths, 
we present density matrix renormalization group (DMRG) \cite{white} 
calculations which allow to reveal the appearance of different phases 
as a function of the deformation parameter $t$. In particular 
we show that local magnetization is able to capture the $t>1$ 
regime where a clear domain wall structure takes place 
independently by the spin value $s$. From the other side 
once $t<1$ the system undergoes to a phase transition of a 
Berezinskii-Kosterlitz-Thouless (BKT) type \cite{bkt} as 
signaled by an exponential opening of the gap. 
Moreover our numerical calculations confirm that for 
this kind of deformation the entanglement entropy is 
bounded and size independent \cite{klich}. Crucially 
we find that this gapped regime can be described solely by a 
non-vanishing value of SOP thus 
showing the topological nature of the $t<1$ deformed Motzkin chains. 
For $s=1$ only one %\blu{string} 
SOP \change{is finite, similarly to what 
happens in the $SU(2)$-Haldane phase for XXZ or AKLT spin-$1$ models,} 
%associated to the spin $SU(2)$ symmetry 
thus revealing the presence of a symmetry-protected 
topological (SPT) order. \change{On the other hand, for $s=2$ different kinds 
of Haldane phases have been obtained \cite{jolicoeur,scalapino}. 
In particular, for the spin $2$ Motzkin chain, we show that 
two SOPs display a constant decay exhibiting a phase similar to
%thus describing the known trivial 
the $SO(5)$-topological Haldane order occuring in $s=2$ AKLT model} 
\cite{tu}. 
Interestingly, unlike the undeformed case $t=1$, 
for $t<1$ the CDP \cite{weinberg95} is valid.
\begin{figure}
\includegraphics[scale=0.55]{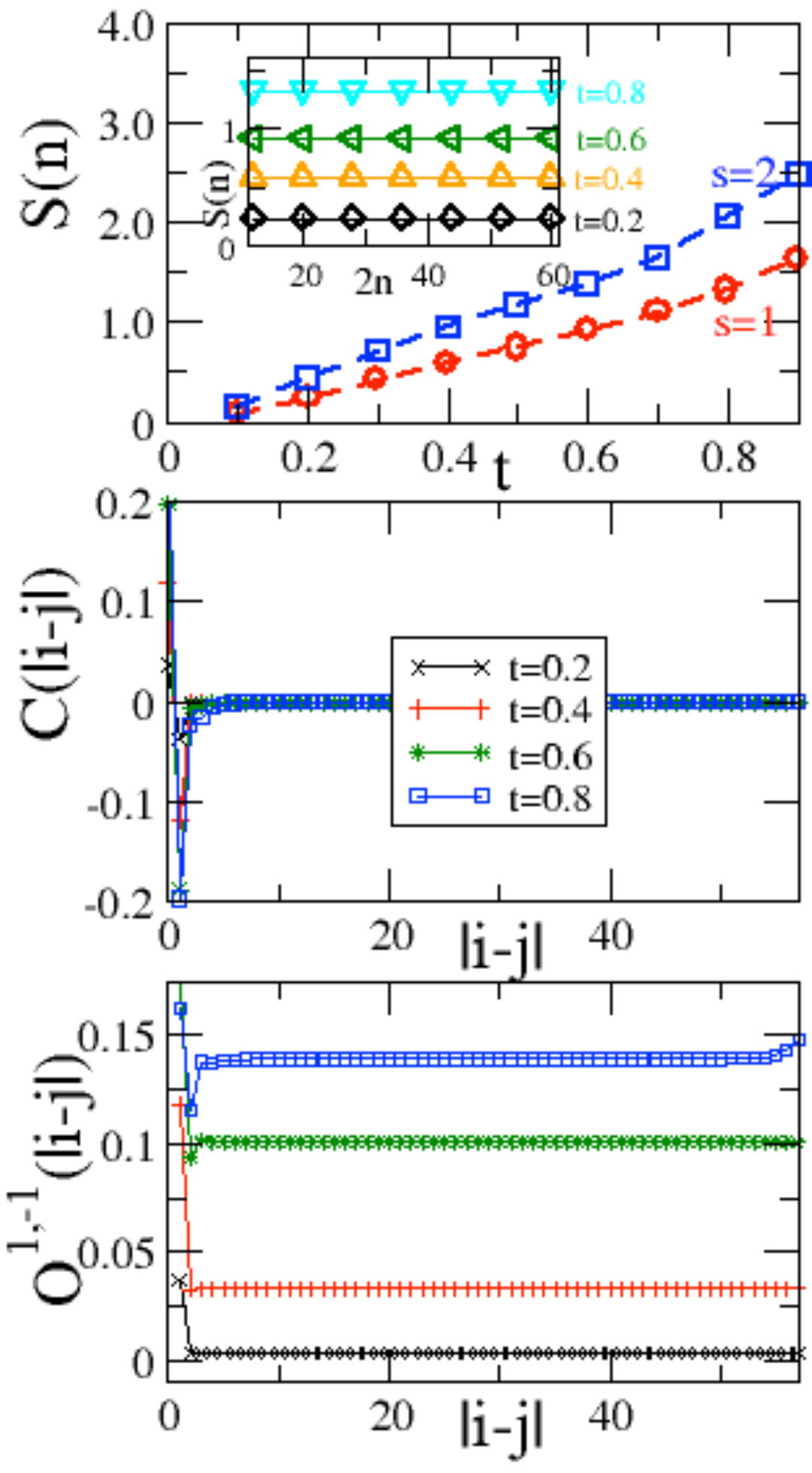}
\caption{(Color online) \textit{a)}: Entanglement entropy $S(A)$ 
for a subsystem having length $n$ with $1\leq i\leq n$ for $s=1$ (red symbols) and 
$s=2$ (blue symbols). The inset shows the constant behavior of 
$S(A)$ as a function of the size $2n$. \textit{b)}: 
$C(|i-j|)$ for different $t<1$ values. \textit{c)}: 
$O^{1,-1}(|i-j|)$ for different $t<1$ values. 
The correlations in the panels $b)$ and $c)$ are evaluated in a 
system of size $2n=60$ with $i$ pinned in the first chain site. 
We checked that different $i$-values do not alter the physical behavior 
of the correlations.}
\label{fig2}
\end{figure}
\paragraph{Model.}
The \change{spin} model we consider \change{has the peculiarity of having a ground state which can be expressed in terms of} Motzkin paths describing all 
the possible $2n$ moves that one can make to go from a point of height 
$h=0$ to an other point of the same $h$ without crossing the $0$ line 
\cite{bravyi12,movassagh14}. As shown in Fig. \ref{fig1} 
\textit{a)} and \textit{b)}, 
\change{spins %the possible moves 
can be seen as moves %spins 
by imposing %the correspondence 
that up/zero/down spin 
%one step 
corresponds to increasing/conserving/decreasing the height of the path. 
%corresponds to create an up/zero/down spin. 
Of course, 
for spin $s=1$ 
%whether 
only uncolored steps (uncolored Motzkin chain) are allowed, 
while larger values of $s$ can be achieved when 
colored steps are possible (colored Motzkin chain).}
%the system can be written in terms of spins with larger $s$ values.
The Hamiltonian reads:
\begin{equation}
H=\sum_{j=1}^{2n-1}\Pi_{j,j+1}(s,t)+\Pi_{\partial}(s)+\sum_{j=1}^{2n-1}\Pi_{j,j+1}^{cross}(s)
\label{ham}
\end{equation}
where \change{$\Pi_{j,j+1}(s,t)=\sum_{k=1}^s\big(|\phi(t)^k\rangle\langle\phi(t)^k|_{j,j+1}+
|\psi(t)^k\rangle\langle\psi(t)^k|_{j,j+1}|+|\Theta(t)^k\rangle\langle\Theta(t)^k|_{j,j+1}\big)$ is the bulk term, 
%describe all the possible moves and 
$\Pi_{\partial}(s)=\sum_{k=1}^s|{-k}\rangle \langle -k|_{1}+|k\rangle\langle 
k|_{2n}$ is the boundary term which} makes more favorable \change{for the first spin to point upward, $|k\rangle$, and the last downward, $|{-k}\rangle$}.  
%step has to be the raising $r$ of a spin and the last move has to be the lowering $l$ of a spin. 
The latter term in Eq.~(\ref{ham}) \change{
$\Pi^{cross}_{j,j+1}(s)=\sum_{k\neq k'}=|k,{-k'}\rangle\langle k,{-k'}|$, 
present only for $s>1$}, ensures the color matching of up and down 
spins with the same $h$. The parameter $t$ appearing in 
$\Pi_{j,j+1}(s,t)$ \change{describes path deformations and    
$|\phi(t)^k\rangle=(1+t^2)^{-1/2}(|k,0\rangle-t|0,k\rangle)$, 
$|\psi(t)^k\rangle=(1+t^2)^{-1/2}(|0,{-k}\rangle-t|{-k},0\rangle)$ and 
$|\Theta(t)^k\rangle=(1+t^2)^{-1/2}(|k,{-k}\rangle-t|0,0\rangle)$. 
%The main point to be noticed here is that such a kinds of 
The deformation induced by $t\neq 1$ keeps the model 
%deformation maintain the $t\neq 1$ model 
frustration-free \cite{klich}, and,} 
%Moreover it is important to underline that 
while for $t=1$ we recover the 
undeformed model 
\cite{bravyi12,movassagh14,dellanna}, for $t>1$ ($t<1$) the paths having 
larger (smaller) $h$ are favored \change{in the ground state}. 
Notice that for $t=1$ one can have analytical expressions 
for the magnetizatazion and the $z-z$ correlation functions, which were 
tested against DMRG results in \cite{dellanna}. However, for $t \neq 1$ 
the corresponding results are not available and therefore we will rely on 
DMRG results to have a physical description of the properties of the model.
\paragraph{$t\geq1$ Regime.}
This latter point explains the $t>1$ behavior 
of the local magnetization $\langle S^z_j\rangle$ observed in 
Fig. \ref{fig1} \textit{c)} and \textit{d)} 
for $s=1$ and $s=2$ respectively. Indeed, since $t>1$ 
makes more probable higher paths, in terms of spins 
this corresponds to a domain wall (DW) structure where the 
up and down spins are separated in two different regions 
of equal length $n$ \cite{note} 
and the zero spins are basically absent. 
Relevantly, as shown in Fig. \ref{fig1} \textit{e)} and \textit{f)}, 
this latter regime is gapless ($\Delta=0$) meaning that the 
difference between the ground-state $E_0$ and the first excited state $E_1$ 
energy goes to zero in thermodynamic limit (TDL). 
The aforementioned features allow to find the analogy between 
Eq. \ref{ham} and the XXZ chains for both spin 
$1$ and $2$ \cite{korepin} for strong negative $z$-anisotropies. 
Further similarities can be also noticed for the $t=1$ case 
where a gapless regime is associated to a power-law decay of the correlation function $\langle S_i^+S_j^-\rangle$ 
and, as exactly shown in \cite{dellanna}, 
an exponential decay of
$\langle S_i^zS_j^z\rangle-\langle S_i^z\rangle\langle S_j^z\rangle$ 
\cite{note2} thus resembling the $XY$ phase of XXZ models but, 
with the key feature that both AL decay and CDP are violated for $s=2$. 
\paragraph{$t<1$ Regime.}
From the other side, as already mentioned, a $t<1$ 
deformation minimizes the height of the possible paths. 
This is clearly visible in the $\langle S_j^z\rangle$ behavior shown in 
Fig. \ref{fig1} \textit{c)} and \textit{d)} where an almost totally 
flat local magnetization with $h=0$ is observed. 
Crucially $\langle S_j^z\rangle$ shows also antiparallel 
peaks at the edges of the chain thus 
supporting the possible presence \change{of} edge states. This effect, 
as explained before, is produced by the $\Pi_{\partial}(s)$ term 
in Eq. (\ref{ham}) which has \change{also the role 
%the first and last path but also to 
of breaking} the ground-state degeneracy. 
The almost flat magnetization can explain also the fact that the entanglement entropy,  
$S(A)=-Tr\rho_A\log_2\rho_A$ of a subsystem $A$,  
is bounded and does not depend on neither the chain nor on the partition 
length \cite{klich}, meaning that the AL scaling is fulfilled. 
Indeed, as it is possible to see in Fig. \ref{fig2} \textit{a)}, 
we find that $S(A)$ is constant at fixed $t$ for any $2n$ while 
it grows almost linearly with the deformation strength. 
The latter is easily explained by the fact that for $t<1$ 
the strength of $t$ actually affects %\blu{only} 
mainly the first and 
the last move with flat $\langle S_j^z\rangle=0$ in the bulk. 
Consequently a larger/smaller $t$ will produce a
higher/lower value of $|\langle S_i^z\rangle|$ in the first and last site, as shown in Fig.~\ref{fig1}, 
thus generating more/less entropy, which is however size-independent (see the inset in the top panel of Fig.~\ref{fig2}) because of the flatness of the paths that mainly contribute to the ground state. Notice that, 
as shown in Fig. \ref{fig2} \textit{a)}, this behavior 
holds for any considered $s$ value. 
A less trivial aspect, conjectured in \cite{klich}, 
emerges by looking Fig. \ref{fig1} \textit{e)} and \textit{f)}, 
namely $t<1$ deformations support the presence of finite gap 
in the TDL. As visible in the same figures, 
for both $s=1$ and $s=2$ the gap opens compatibly 
with an exponential decay $\Delta\sim\exp{(-b/\sqrt{t_c-t})}$, 
being $t_c=1$ and $b$ a fitting parameter, thus signaling a BKT-like 
phase transition. 
In integer spin chains gapped regime can 
be usually associated to either antiferromagnetic (AF) order described 
by the two points correlation functions
\begin{equation}
C(|i-j|)=\langle S_i^zS_j^z\rangle-\langle S_i^z\rangle\langle S_j^z\rangle
\label{szsz}
\end{equation}
or to Haldane orders described by a SOP
\change{
\begin{equation}
O^{k,\bar{k}}(|i-j|)=\langle L^{k,\bar{k}}_ie^{\imath\pi\sum_{i\leq \ell<j}L_\ell^{k,\bar{k}}}L^{k,\bar{k}}_j\rangle
\label{str}
\end{equation}
where $L^{k,\bar{k}}=|k\rangle\langle k|-|{-k}\rangle\langle{-k}|$.} 
Notice that, for $s=1$, \change{$k(\bar{k})$} can be solely equal 
to $1(-1)$ thus $L^{1,-1}_i=S^z_i$ while for $s=2$, \change{$k(\bar{k})$} 
can take the values $1(-1)$ and $2(-2)$ and 
\change{$S_i^{z}=2\,L^{2,-2}_i+L^{1,-1}_i$.}  
%the previous equivalence no longer holds. 
The important informations encoded in such non-local order parameters 
is that, once it is finite, Eq.~(\ref{str}) describes a 
topological phase, usually called HP, 
with hidden antiferromagnetism (HAF). The HAF order is given by the 
fact that it can not be described by usual two point correlation functions 
Eq. (\ref{szsz}) thus describing a phase where spins up and down 
are rigorously alternated and separated by a random number of zero spins. 
Of course, while for $s=1$ the HP can be given only by alternating $+1$ and 
$-1$ spins thus signaled by a finite $O^{1,-1}(|i-j|)$, for $s=2$ 
the hidden order can be signaled, as it will be clear in the following, 
by two or even solely one finite $O^{k,\bar{k}}(|i-j|)$. 
\begin{figure}
\includegraphics[scale=0.55]{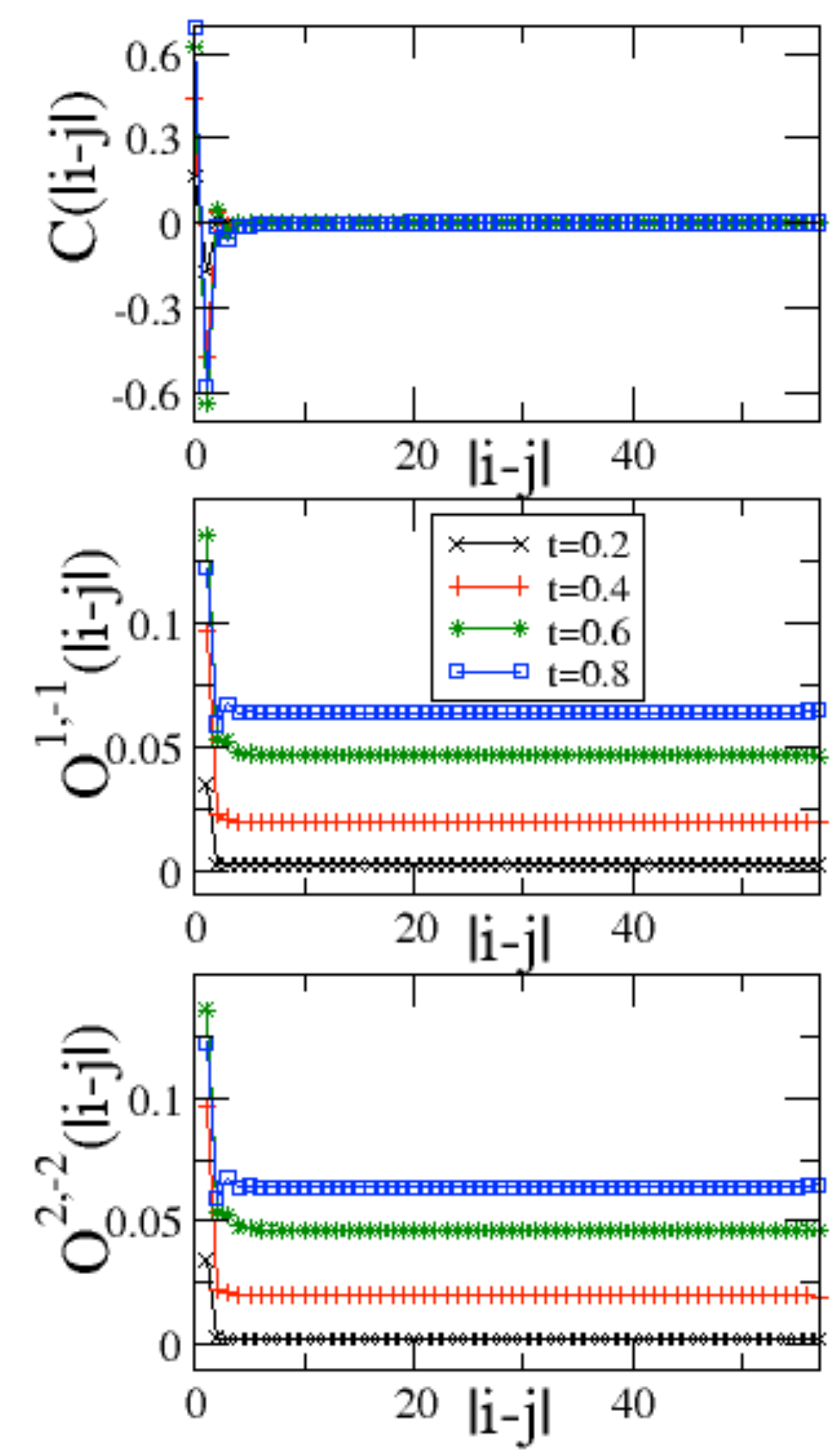}
\caption{(Color online) \textit{a)}: $C(|i-j|)$ 
for different $t<1$ values. \textit{b)}: $O^{1,-1}(|i-j|)$ 
for different $t<1$ values. \textit{c)}: $O^{2,-2}(|i-j|)$ 
for different $t<1$ values. All the correlation are 
evaluated in a system of size $2n=60$ with $i$ pinned in the first 
chain site. We checked that using different values 
of $i$ do not alter the physical behavior of the correlations.}
\label{fig3}
\end{figure}
\subparagraph{$s=1$ case.} 
Here we start our analysis with the $s=1$ case by evaluating both 
$O^{1,-1}(|i-j|)$ and $C(|i-j|)$ for different $t<1$ values. 
Fig. \ref{fig2} clearly shows that while $C(|i-j|)$ 
rapidly decays to zero the SOP 
remains constant as a function of the distance 
thus signaling the presence of a HP. This aspect, 
in analogy with XXZ chains, supports our prediction 
regarding the BKT nature of the phase transition. 
We also checked that the %\blu{strings} 
SOPs along transverse directions decay. 
Furthermore, as visible \change{in Fig.~\ref{fig2} {\it c)}}, 
$O^{1,-1}(|i-j|)$ saturates 
to a constant value which becomes \change{bigger the larger is $t$}. 
At a first look this aspect could seem counterintuitive 
since one expects that the larger is the gap, the 
stronger is the %\blu{string order} 
SOP. Nevertheless an easy interpretation 
of the $O^{1,-1}(|i-j|)$ behavior as function of $t$ comes 
by the geometrical meaning of deformations. Indeed, as explained before, 
a small $t$ favors the paths with low $h$. 
Intuitively, one can argue that the path with smaller $h$ 
is the one where the first and last move corresponds to respectively 
the rising and the lowering \change{steps} with in the middle a 
series of flat moves. 
This means that the number of $+1$, $-1$ spins producing the
HAF order is minimized by \change{reducing} $t$, 
thus producing a lower saturation value of the %\blu{string order} 
SOP. 
Nevertheless we checked that even very small deformations support 
the presence of a constant $O^{1,-1}(|i-j|)$, \change{suddenly} 
disappearing ($O^{1,-1}(|i-j|)=0$) for $t=1$, 
thus allowing to unambiguously conclude that the uncolored 
$t<1$ Motzkin chain has topological order 
with HAF. 
%Due to the spin symmetry, 
\change{This} phase is usually called $SU(2)$-Haldane phase and it has been 
observed both in spin-$1$ XXZ \cite{denNijs,kennedy} chains and 
in the AKLT model \cite{kennedy1}. 
\change{We will keep this nomenclature even if for our model only the operator 
$\sum_{i}S^z_i=\sum_iL_i^{1,{-1}}$ commutes with the Hamiltonian. 
This is similar to what happens in the spin-$1$ XXZ model when a single 
ion anisotropy term is included, breaking the $SU(2)$ invariance but 
preserving the HP.}
For certain systems %\blu{with conformal invariance} 
like the spin-$1$ XXZ model, the topological order is also
captured by an even degeneracy of the entanglement spectrum 
(ES) \cite{pollmann}. From the other side, 
%\blu{when Hamiltonians cannot be described by conformal field theory (CFT),} 
in different models,  
like for instance the AKLT \cite{fan} or exotic bosonic Hamiltonians 
\cite{batrouni}, the ES does not present even degeneracy 
but the topological order is assured by the presence of edge modes 
and finite %\blu{strings} 
SOPs. We checked that 
%\blu{due to the lack of a possible description in terms of CFT \cite{movassagh14}} 
this happens also in our case where the ES of $t<1$ 
deformed Motzkin chains does not display any degeneracy. 
Nevertheless edge modes, visible in Fig. \ref{fig1} \textit{c)} and 
\textit{d)}, and the finite %\blu{string} 
SOP in Fig \ref{fig2} 
assure the topological order \cite{wenPRB2011}. The latter has a further fundamental 
property due to fact that it appears for an odd value $s$ of the spin. 
Indeed, once a HP takes place for odd spins $s$, 
SPT order \cite{wen1} is generated. 
This is given by the fact that the edge modes fractionalize 
in two half-integer spins which cannot be removed unless 
in presence of a phase transition or an explicitly symmetries breaking. 
This consideration allows to conclude that the $s=1$ \change{version of} 
Eq.~(\ref{ham}) with $t<1$ deformations supports the presence of SPT 
topological order with bounded and size independent entanglement 
thus strongly characterizing the the Motzkin chains.
\subparagraph{$s=2$ case.}
As shown in \cite{movassagh14,dellanna}, 
the $s>1$ undeformed $t=1$ chains have much more intriguing 
properties with respect to lower spin case. These are induced by the 
presence of colors which increase the symmetry 
of the system. 
As for the $s=1$, $s=2$ XXZ Heisenberg \change{and AKLT models can} 
support the presence of gapped phases for positive $z$-anisotropies. 
The gap can again be associated to AF order detected by $C(|i-j|)$ 
\change{or to different kinds of %the $SO(5)$ 
HP, see for instance \cite{jolicoeur,orus} 
and reference therein. In particular in such systems 
SPT topological order} is signaled by finite values of 
both $O^{1,-1}(|i-j|)$ and $O^{2,-2}(|i-j|)$. 
%Notice that the latter is the case 
%also of the $s=2$ AKLT model. 
%Crucially the $SO(5)$-HP strongly differs from the $SU(2)$ case since for 
%even $s$ the edge modes remain integer thus the ground-state 
%can be adiabatically turned into a trivial state 
%without any symmetry breaking. As consequence the SPT regime 
%of the $s=1$ case is substituted by a trivial topological order. 
%Nevertheless, 
\change{Moreover}, as conjectured in \cite{oshikawa} 
and shown in \cite{orus,aschauer}, single ion anisotropy 
terms can support the formation of a SPT $SU(2)$-Haldane order even 
for $s=2$. Our calculations in Fig. \ref{fig1} \textit{f)} show 
that again the colored Motzkin chain is gapped for $t<1$ and 
the gap is associated to HAF since $C(|i-j|)$ 
has a clear exponential decay rapidly saturating to $0$ as shown 
in Fig. \ref{fig3} \textit{a)}. From the other side both 
$O^{1,-1}(|i-j|)$ and $O^{2,-2}(|i-j|)$ have a constant and basically 
equal behavior thus clarifying that the $s=2$ Motzkin chains 
with $t<1$ deformations support the presence of an 
$SO(5)$-HP. 
\change{It is worth stressing that $SO(5)$ is not the symmetry of our model, rather $U(1)\times U(1)\times \mathbb{Z}_2$, since only $\sum_iL_i^{2,{-2}}$ and $\sum_i L_i^{1,{-1}}$ commute with the Hamiltonian, 
like in the s=2 AKLT model when the term $\sum_i(S^z_i)^2$ is switched on. Also in that case $SO(5)$-Haldane phase survives once the symmetry is lowered from $SO(5)$ to $U(1)\times U(1)$ \cite{orus2}. 
In our case the symmetry is supplemented by the invariace under interchanging the two colors ($\mathbb{Z}_2$). This is the reason why $O^{1,{-1}}(|i-j|)$ and $O^{2,{-2}}(|i-j|)$ are the same, as shown in Fig.~\ref{fig3}.}
Moreover it is important to notice that, 
in analogy with the $s=1$ case, \change{the SOPs} %$SO(5)$-Haldane oder 
become stronger by increasing $t$. Fig. \ref{fig3} \textit{b)} 
also shows a further information encoded in the $C(|i-j|)$ behavior. 
Indeed, on the contrary to the $t=1$ regime \cite{dellanna}, 
its exponential decay is associated to a 0 edge-to-edge value thus holding 
\change{the CDP}. 
%Thus it allows to conclude how 
\change{The} opening of a 
gap in a colored Motzkin chain, \change{therefore,} restores the pure locality of the model in Eq. (\ref{ham}), in agreement with the general findings for 
gapped local Hamiltonians \cite{hastings}.
\paragraph{Conclusions and Perspectives.}
In conclusion, our results demonstrate the existence of topological Haldane orders in a 
new class of spin Hamiltonians. Furthermore we have shown the behavior of 
Motzkin chains as a function 
of the deformation strength $t$. While the undeformed $t=1$ case has 
$XY$-like features, for $t>1$ the system presents a gapless domain wall structure. 
From the other side, at $t=1$, we presented evidences of a BKT-like phase transition, 
characterized by an exponential opening of the gap, occurring for any $t<1$ values. The gapped regime is associated to SPT  
hidden Haldane antiferromagnetic orders signaled by finite values of 
string non-local order parameters. 
%While for $s=1$ the Haldane phase has symmetry-protected topological order with fractionalized edge modes, $s=2$ chains have trivial $SO(5)$ topological features. For this reason it would very interesting to look for single 
%ion anisotropy terms-like conserving the Motzkin paths constrains 
%but making possible the appearance of a symmetry-protected Haldane regime 
%in the $s=2$ chain. 
The two possible Haldane orders have the peculiarity 
\change{of having} an entanglement entropy independent from both block and chain size.
Moreover our results suggest that it would be very interesting 
to have a physical implementation of the Motzkin spin chains. 
In this respect cold atomic systems, which 
have been already proposed to simulate several kinds of spin Hamiltonians 
with topological orders \cite{dallatorre}, could provide a possible 
physical platform to implement Motzkin chains. Their experimental 
realization could be relevant for technological achievements since, 
from one side, symmetry-protected topological orders have been proposed as ideal candidates towards the realization quantum devices like quantum 
repeaters \cite{verstraete} and substrate for measurement-based 
quantum computation \cite{else} while, from the other, side 
Motzkin paths may 
have applications in the field of polymer absorption \cite{rensburg}.
Finally we underline that, in future works, it would 
be very interesting to study the gapped regimes in the fermionic version of the $s=3/2$ Fredkin 
model where exotic Haldane regimes can take place \cite{fazzini}.
\begin{acknowledgments}
{\it Acknowledgments:} Discussions with D.K. Campbell 
and O. Salberger are acknowledged. A. Montorsi and E. Orlandini are warmly are acknowledged for suggesting us crucial references.
\change{Useful correspondence with H.-H. Tu is also kindly acknowledged.}
We acknowledge the support and hospitality of 
the Simons Center for for Geometry and Physics in Stony Brook where this 
paper has been conceived during the program 
{\em Entanglement and Dynamical Systems}. 
LD thanks SISSA and LB thanks University of Padova 
for kind hospitality. LD acknowledges 
financial support from MIUR through FIRB Project  No. RBFR12NLNA\_002. LB and 
AT acknowledge support from the European STREP MatterWave. LB also acknowledges ERC Starting Grant TopoCold for financial support. Finally, 
LB and AT acknowledge Karma cluster in Trieste for CPU time. \\
{\it Note added:} After the submission of this paper, two 
papers appeared on arXiv where a Motzkin spin chain is considered 
introducing a field theory approach to study certain 
observables and entanglement measures \cite{arXiv}. 
We think that addressing with such approach 
the properties discussed in the present paper 
would be very interesting, in particular the discussion on the BKT nature 
of the transition at $t=1$.
\end{acknowledgments}


\begin{thebibliography}{99}
\bibitem{note} The fact the spins up and down 
are separated in two regions of equal length is due to the fact that 
we are imposing a vanishing total magnetization 
$S^z=\sum_{j=1}^{2n}S_j^z=0$.
\bibitem{note2} For $s=2$ the exponential decay is verified if the 
$i$-index in the correlation is sufficiently far from the chain edge, 
while for $s=1$ the exponential decay is always fulfilled.
\end{thebibliography}
\end{document}